\documentclass[preprint,12pt]{elsarticle}

\usepackage{amssymb}
 \usepackage{amsthm}
 \usepackage{lineno}
\usepackage{amssymb}
\usepackage{graphicx}
\usepackage{times}
\usepackage{multirow}
\usepackage{bm}
\usepackage{graphics}
\usepackage{subfigure}
\usepackage{algorithm}
\usepackage{ifpdf}
\usepackage{booktabs}
\usepackage{graphicx,natbib,amssymb,lineno,graphics,subfigure}

\ifpdf
\usepackage[%
  pdftitle={Instructions for use of the document class
    elsart},%
  pdfauthor={Simon Pepping},%
  pdfsubject={The preprint document class elsart},%
  pdfkeywords={instructions for use, elsart, document class},%
  pdfstartview=FitH,%
  bookmarks=true,%
  bookmarksopen=true,%
  breaklinks=true,%
  colorlinks=true,%
  linkcolor=blue,anchorcolor=blue,%
  citecolor=blue,filecolor=blue,%
  menucolor=blue,pagecolor=blue,%
  urlcolor=blue]{hyperref}
\else
\usepackage[%
  breaklinks=true,%
  colorlinks=true,%
  linkcolor=blue,anchorcolor=blue,%
  citecolor=blue,filecolor=blue,%
  menucolor=blue,pagecolor=blue,%
  urlcolor=blue]{hyperref}
\fi

\makeatletter
\def\elsartstyle{%
    \def\normalsize{\@setfontsize\normalsize\@xiipt{14.5}}
    \def\small{\@setfontsize\small\@xipt{13.6}}
    \let\footnotesize=\small
    \def\large{\@setfontsize\large\@xivpt{18}}
    \def\Large{\@setfontsize\Large\@xviipt{22}}
    \skip\@mpfootins = 18\p@ \@plus 2\p@
    \normalsize
}
\@ifundefined{square}{}{}
\makeatother

\def\bbbr{{\rm I\!R}}

\def\Skip{\par\bigskip\nobreak}

\newtheorem{definition}{Definition}
\newtheorem{proposition}{Proposition}
\def\Skip{\par\bigskip\nobreak}
\def\dj{d\kern-0.4em\char"16\kern-0.1em}
\def\Dj{\mbox{\raise0.3ex\hbox{-}\kern-0.4em D}}

\newcommand{\thb}{\bm{\theta}}
\newcommand{\phib}{\bm{\phi}}

\newcommand{\yb}{\mathbf{y}}

\newcommand{\Kb}{\mathbf{K}}
\newcommand{\Ib}{\mathbf{I}}

\newcommand{\alphab}{\bm{\alpha}}
\newcommand{\thetab}{\bm{\theta}}
\newcommand{\betab}{\bm{\beta}}

\journal{Pattern Recognition Letters}

\begin{document}

\begin{frontmatter}


 \author{Javier Gonz\'alez\footnote{j.gonzalez.hernandez@rug.nl}, Ivan Vuja\v{c}i\'c and Ernst Wit}
\address{Johann Bernoulli Institute of Mathematics and Computer Science,\\ University of Groningen, Nijenborgh 9. 9747 AG Groningen,  The Netherlands\\ 
}


\title{\textbf{Reproducing kernel Hilbert space based estimation of systems of ordinary differential equations}}


\begin{abstract}
Non-linear systems of  differential  equations have attracted the
interest in fields like system biology, ecology or biochemistry, due to their flexibility and their ability to describe
dynamical systems. Despite the importance of such models in many
branches of science they have not been the focus of systematic statistical
analysis until recently. In this work we propose a general
approach to estimate the parameters of systems of differential equations measured with noise. Our methodology is based on the maximization of the penalized likelihood 
where the system of differential equations is used as a penalty. To do so, we use a Reproducing Kernel Hilbert Space approach that allows us  to formulate the estimation problem as an unconstrained numeric maximization problem easy to solve. The proposed method is tested with synthetically simulated data and it is used to estimate the unobserved transcription factor CdaR in \emph{Steptomyes coelicolor} using gene expression data of the genes it regulates.
\end{abstract}


\end{frontmatter}


\section{Introduction}

Despite the fact that differential equations are a common
modelling tool within science and engineering, statistical methods
for estimating such models have only received wide-spread attention during the last few years. The
difficulty of solving differential equations in general has been a major
stumbling block for efficient statistical procedures. Only in the last six
years, serious progress has been made on the estimation of parameters
within systems of differential equations measured with noise. More importantly,
solving the differential equation in these methods is not necessary for estimating
the parameters of the differential equation. \citet{Ramsay07}
introduced an approximate, two-stage maximum likelihood estimation procedure,
where the solution of the differential equation was linked to a
smoothed version of the data. In \citep{Calderhead08} a
Bayesian method is proposed similar in spirit, whereby the solution of the ordinary differential equation (ODE)
was given as a Gaussian process prior. Estimation was effectively again
a two-stage process, where the \emph{product of experts} provided a
crucial link. In \citep{Steinke08} a kernel method is developed for
estimating 1-dimensional, periodic differential equations. In \citep{2013arXiv1306.2365C} a fully Bayesian inferential framework is developed to quantify uncertainty in ODE models. In this paper, we combine the frequentist set-up, such as in \citep{Ramsay07}, with the kernel approach of
\citep{Steinke08}. The main advantage is that we can define the estimation problem explicitly as a maximum likelihood problem, whereby the differential equation is interpreted as a constraint. 
By introducing a reproducing kernel Hilbert space (RKHS), we transform the constrained maximization problem into an unconstrained maximization problem.  We detail this idea in  Sections 3 and 4 after 
a revision of the main properties of RKHSs and penalized likelihood models in Section \ref{sec:prelim}. In Section 5 we focus on the implementation of our methodology. Sections 6 and 7 illustrate the
behaviour of the technique in simulated and real data scenarios, respectively. We conclude in Section 8 with a discussion of practical results of this work.

\section{The RKHS framework: Green's functions, penalized likelihood models and Gaussian Processes}\label{sec:prelim}

Reproducing Kernel Hilbert Spaces \cite{bib:aroszajn50,bib:cucker2001} have played an important role in a number of successful applications in the last decades \cite{Parzen1970,bib:wahba90,bib:moguerza06,Hofmann2008}. In this work we use this theory for ODE estimation in the context of constrained likelihood models.

An RKHS is a Hilbert space of functions where all the linear evaluation functionals ${\cal F}_t : \mathcal{H} \rightarrow \bbbr$ such that ${\cal F}_t (x) = x(t)$, for ${t} \in T$, are bounded. By virtue of the Riesz representation theorem, for each ${t} \in T$ there exists $k_t \in \mathcal{H}$ such that for every $x \in \mathcal{H}$, $x(t) = \langle k_t , x \rangle $, where $\langle,\rangle$
denotes the inner product in $\mathcal{H}$. The RKHS $\mathcal{H}$ is uniquely characterized by a continuous, symmetric and positive definite function $K : X \times X \rightarrow \bbbr$
named Mercer Kernel or reproducing kernel for $\mathcal{H}$ \citep{bib:aroszajn50}.
All linear combinations $x(t) = \sum_i\alpha_iK(t,t_i)$ where $\alpha_i\in \bbbr$ and $t_i \in T$ equipped with the adequate inner product form a space which is dense in $\mathcal{H}$. See \cite{bib:wahba90,bib:cucker2001} for details.

 
Let $T$ be a closed interval and consider a random sample $S=\{ (y_i,t_i)\in \bbbr\times T \}_{i=1}^n$ made up of $n$ independent observations. Assume that the conditional distribution of $y_i$ given $t_i$ is Gaussian such that $y_i | t_i, x,\sigma^2 \sim \mathcal{N}(x(t_i),\sigma^2),$
where $\sigma^2$ is the variance of the model and $x: T \rightarrow \bbbr$ is the target regression function to be estimated. To estimate $x$, one needs to restrict $x$ to belong to a particular class of models, which can be done by penalizing the likelihood with a convex functional acting on $x$, generally a norm or seminorm in some Hilbert space $\mathcal{H}$. A widely studied approach is to use a differential operator $P$ to impose smoothing conditions on $x$ \cite{Green_Silverman_1994}. In the Gaussian case, assuming that $\sigma^2$ is known,  the resulting penalized log-likelihood is 
\begin{equation}\label{eq:penalized.likelihood}
l_{\lambda}(x|S, \sigma^2) = -\frac{1}{2\sigma^2}\|x(\textbf{t})-\textbf{y}\|^2 - \frac{\lambda}{2}\|Px\|^2_{L_2},
\end{equation}
where $\|\cdot\|_{L_2}$ is $L_2$ norm, $\textbf{t}=(t_1,\dots,t_n)$ , $\textbf{y}=(y_1,\dots,y_n)^T$ and $\lambda>0$ controls the balance between the fitting to the data and the smoothness of the model.

The maximization of (\ref{eq:penalized.likelihood}) can be written as a regularization problem in an RKHS \cite{bib:Berlinet2004,58326}. To do so, a way to connect the differential operator $P$ with a kernel in a Hilbert space is required. A semi-constructive way to build such kernel is to use the concept of the Green's function. The Green's function of a linear operator $P$ is a function $\mathcal{G}: T \times T \rightarrow \bbbr$ that satisfies $P\mathcal{G}(t,z) = \delta(t-z)$ for $\delta$ the Dirac distribution. Roughly speaking $\mathcal{G}$ is the inverse of $P$ \cite{duffy2001}. Consider a differential operator $P$ and let be $\mathcal{K}$ a Green's function of $P^{\ast}P$ where $P^{\ast}$ is the adjoint operator of $P$. It can be shown \cite{bib:aroszajn51} that 
$\|Px\|^2_{L^2}=\|x\|^2_{\mathcal{H}_{\mathcal{K}}},$ for $\|\cdot\|_{\mathcal{H}_\mathcal{K}}$ the norm in the RKHS defined by $\mathcal{K}$. 

Expression (\ref{eq:penalized.likelihood}) can be understood as a projection mechanism onto the finite dimensional space spanned by $\{\mathcal{K}(t,t_i) \}$ when the penalization term $\|Px\|^2_{L^2}$  is replaced by $\|x\|_\mathcal{K}^2$ \cite{Kimeldorf1970, Ma_Dai_Klein_Klein_Lee_Wahba_2010}. In particular, the maximizer of (\ref{eq:penalized.likelihood}) is the function $\hat{x}(t) = \sum_{i=1}^n \hat{\alpha}_i \mathcal{K}(t,t_i)$ with coefficients $\hat{\alpha}_i$, obtained by maximizing 
\begin{equation}\label{eq:reg2}
l_{\lambda}(x| S, \sigma^2) = -\frac{1}{2\sigma^2} \|\textbf{y}-\mathcal{K}[\textbf{t}]\alphab\|^2-\frac{\lambda}{2}\alphab^T\mathcal{K}[\textbf{t}]\alphab,
\end{equation}
with respect to $\alphab = (\alpha_1\dots,\alpha_n)^T$, where $\mathcal{K}[\textbf{t}]$ is the matrix with entries $(\mathcal{K}[\textbf{t}])_{ij}=\mathcal{K}(t_i,t_j)$. By using standard methods of differential calculus it can be shown that the solution to the maximization problem in (\ref{eq:reg2}) is given by $\hat{\alphab}= (\lambda 
\sigma^2\textbf{I}_{n} + \mathcal{K}[\textbf{t}])^{-1}\textbf{y},$ where $\textbf{I}_n$ is the $n$-dimensional identity matrix \citep{bib:vapnik95,Hofmann2008}.


\section{Explicit ODE and regularization approach}
\subsection{Explicit ODE}
In this work we are interested in dynamical systems with $m$ interacting elements evolving in some closed time interval $T$. We denote by $x_j:T \rightarrow \bbbr$ for $j=1,\dots,m$, the functions describing the evolution of the elements of the system and by $u_j:T \rightarrow \bbbr$ for $i=1,\dots,p$, the action of $p$ external forces.  In compact notation, we denote by $\textbf{x}(t)=(x_1(t),\dots,x_m(t))^T$ and $\textbf{u}(t)=(u_1(t),\dots,u_p(t))^T$ the vectors of state variables and external forces respectively. We assume that each state variable $x_j$ satisfies   
\begin{equation}\label{eq:system.diff}
P_{\thb_j} x_j(t)= f_{j}(\textbf{x}(t),\textbf{u}(t),\betab), \,\,\, j=1,\dots,m, \,\,\, t \in T, 
\end{equation}
where $P_{\thetab_j} = \sum_{k=1}^d \theta_{jk}d^{k-1}/dt$ is the linear differential operator associated to the $j$-th equation of the system, 
which is defined by parameters $\thetab_j=(\theta_{j1},\dots,\theta_{jd})^T$, and $f_{j}$ is a known function depending on $t$ through $\textbf{x}(t)$ and $\textbf{u}(t)$ for a vector of parameters $\betab$. In the sequel, we refer to the whole set parameters of the system by $\{\thetab,\betab\}$, where $\thetab= \{\thetab_1,\dots,\thetab_m\}$. 

Typically, a finite sample of the states $\textbf{x}$ is observed at a grid of time points $\textbf{t} = (t_1,\dots,t_n)^T$, that for simplicity we assume equal for all the states. The sample $\textbf{y}(\textbf{t})$ is made up of noisy measurement of the states of the ODE. That is, $\textbf{y}(\textbf{t})$ satisfies 
$$\textbf{y}(\textbf{t})= \textbf{x}(\textbf{t}) + \epsilon(\textbf{t}),$$ 
where $\epsilon(\textbf{t})$ represents a noise process. We assume $\epsilon(\textbf{t})$ is independent multivariate zero-mean Gaussian noise with variance $\sigma_j^2$ for each $j$th state. Other noise models are possible though not trivial.


Let $\textbf{y}_j$ indicate the available data for state $j$ and let $x_j(\textbf{t})$ indicate the vector of values
corresponding to evaluated $j$-th state at time points $\textbf{t}$. The \emph{log-likelihood} of the model given the sample $\textbf{y}(\textbf{t})$ is given by
\begin{equation}\label{def:like.univ.ode}
l(\thetab,\betab, \textbf{x}| \textbf{y}(\textbf{t}))=- \sum\limits_{j=1}^{m}\frac{1}{2\sigma^2_j}\|\textbf{y}_j-x_j(\textbf{t})\|^2 \,\, \mbox{given that}\,\,  P_{\thb_j} x_j(t)= f_{j}(\textbf{x}(t),\textbf{u}(t),\betab),
\end{equation}
where $\textbf{x}$ depends on $\{\thetab,\betab\}$ although it is not explicitly specified. Indeed, for each set of parameters $\{\thetab,\betab\}$ a family of feasible solutions of the system of differential equation
 is available for different initial conditions $\textbf{x}(0)$.

\subsection{Regularization }
The maximum likelihood estimators of $\{\betab, \thetab\}$ and $\textbf{x}$ require explicit solutions of the differential equation, which are generally unknown. 
Alternatively, computational ODE solvers result intractable in cases with a large number of parameters. Our goal is to reformulate (\ref{def:like.univ.ode}) in order to obtain a computationally tractable 
solution that does not require an explicit solution of the differential equation. The key element of our approach is the penalized log-likelihood 
\begin{equation}\label{def:like.univ.ode.pen}
l_{\lambda}(\thetab,\betab, \textbf{x}| \textbf{y}(\textbf{t}))=- \sum\limits_{j=1}^{m}\frac{1}{2\sigma^2_j}\|\textbf{y}_j-x_j(\textbf{t})\|^2 - \frac{\lambda}{2} \sum\limits_{j=1}^{m} \Omega(x_j)
\end{equation}
where $\Omega(x_j)$ is a convex functional that adds to the probabilistic model of the data the information provided by the ODE and $\lambda>0$ is a regularization parameter. As we detail next, an RKHS representation of the ODE can be used to define $\Omega(x_j)$ avoiding the computational burden of numerical ODE solvers.

\section{RKHS representation of ODE systems}
If the system is homogeneous, each equation 
\begin{equation}\label{eq:homogeneous}
P_{\thb_j} x_j= 0,
\end{equation}
is independent of the rest. Consider the set of functions $\mathcal{H}_j=\{x: \|P_{\thb_j}x_j\|^2=0 \}$. By the definition of the norm, $\mathcal{H}_j$ contains all the solutions of the differential equation $P_{\thb_j} x_j= 0$ for some fixed $\thetab_j$: $\|P_{\thb_j}x_j\|^2=0$ if and only if $P_{\thb_j}x=0$. Using the connection between differential operators and RKHS detailed in Section \ref{sec:prelim}, we can express the set $\mathcal{H}_j$ as $\mathcal{H}_j=\{x:\|x \|^2_{\mathcal{K}_{\thetab_j}}=0 \}$ where $\mathcal{K}_{\thb_j}$ is the Green's function of $P^{\ast}_{\thb_j}P_{\thb_j}$ and $\| \cdot\|_{\mathcal{K}_{\thb_j}}$ the norm in the RKHS defined by $\mathcal{K}_{\thb}$. Therefore, it makes sense to define
\begin{equation}\label{eq:norm}
\Omega(x_j) = \|P_{\theta_j}x_j\|= \|x_j \|^2_{\mathcal{K}_{\thb_j}},
\end{equation}
as penalty and proceed using the properties of penalized likelihood models in RKHS described in Section 2. Note that by using this penalty in (\ref{def:like.univ.ode.pen}) each $x_j$ is assumed to be an element in $\mathcal{H}_{\thb_j}$ rather than a solution of the differential equation unless we force $\| x\|_{\mathcal{K}_{\thb_j}}^2$ to be zero, which can be imposed by taking $\lambda \rightarrow \infty$. In this case, approaches in (\ref{def:like.univ.ode}) and (\ref{def:like.univ.ode.pen}) are equivalent. 

The main drawback is that a closed-form expression for the Green's function of $P^{\ast}_{\thb_j}P_{\thb_j}$ is rarely available. However, note that as shown in (\ref{eq:reg2}), to solve the regularization problem only the evaluation of $\mathcal{K}_{\thetab_j}$ on the sampled time points is required. Consider the difference equation given by
\begin{equation}\label{eq:diif.eq}
\textbf{P}_{\thb_j}\textbf{x}_j= 0,
\end{equation}
where $\textbf{P}_{\thb_j}= \sum_{k=1}^d\theta_{jk} \textbf{D}^{k-1}$ is a d-order polynomial difference operator (matrix) acting on elements defined on $\textbf{t}$ and where $\textbf{D} \in \bbbr^{n\times n}$ is a first order difference operator such that $(\textbf{D}\textbf{x}_j)_1 = (t_2-t_1)^{-1}(\textbf{x}_{j,2}-\textbf{x}_{j,1})$,  $(\textbf{D}\textbf{x}_j)_i = (2(t_{i+1}-t_{i-1}))^{-1}(\textbf{x}_{j,i+1}-\textbf{x}_{j,i-1})$ and $(\textbf{D}\textbf{x}_j)_n = (t_n-t_{n-1})^{-1}(\textbf{x}_{j,n}-\textbf{x}_{j,n-1})$. Note that this this difference operator is different to the one proposed in \citep{Steinke08}, which it is only valid for periodic differential equations. The following proposition holds.
\begin{proposition}\label{theo:discretegreen}
\citep{Steinke08} Denote by $\mathcal{H}$ the space of functions (vectors) $\textbf{x}_j: \textbf{t} \subset T \rightarrow \bbbr$ equipped with the usual $L_2$ inner product $\langle\cdot,\cdot\rangle$.
Then the Green's function of the difference operator $\textbf{P}_{\thb_j}^{\ast}\textbf{P}_{\thb_j}$, where $\textbf{P}_{\thb_j}^{\ast}$ is the adjoint (transpose) operator of $\textbf{P}_{\thb_j}$, is the $s \times s$ dimensional matrix
\begin{equation}\label{eq:K}
\textbf{K}_{\thb_j}=(\textbf{P}_{\thb_j}^{T}\textbf{P}_{\thb_j})^{-1}.
\end{equation}
\end{proposition}
Roughly, the proof uses the fact that $\bbbr^n$ is an RKHS whose reproducing kernel is the location function $\delta_{i}$, whose inner product with $x_j(\textbf{t})$ yields $\delta_i^Tx_j(\textbf{t}) = x_j(t_i)$ and therefore only the inverse of  $\textbf{P}_{\thb_j}^{T}\textbf{P}_{\thb_j}$ in needed. See \citep{Steinke08} for details.

Penalty $\Omega(x_j)$, although originally defined for the function $x_j$, affects the regularization process through the values of $x_j$ evaluated at a finite dimensional grid, generally the one corresponding to the available sample. By replacing the differential equation in (\ref{eq:homogeneous}) by the difference equation (\ref{eq:diif.eq}) and using Proposition 1, we define the penalty of the log-likelihood as
$$ \Omega(x_j) = \|\textbf{P}_{\thetab_j}\textbf{x}_j\|^2= \textbf{x}_j^T \textbf{P}_{\thetab_j}^T\textbf{P}_{\theta_j} \textbf{x}_j = \alphab_j^T\textbf{K}_{\thb_j}\alphab_j, $$
where $\alphab_j = \textbf{K}_{\thb_j}^{-1} \textbf{x}_j$. 

\subsection{Generalizations to nonhomogeneous equations}
In general, the interest in inferring parameters of ODE is for systems which are not homogeneous. In the same spirit of above, one might like to consider
\begin{equation}\label{pen2}
\Omega(x_j) = \| P_{\thb_j} x_j - f_{j}(\textbf{x},\textbf{u},\betab)\|^2,
\end{equation}
as a penalty. However (\ref{pen2}) cannot be directly used as penalizer for two reasons. Expression (\ref{pen2}) cannot be reformulated as a norm of $x_j$ in an RKHS. Note that when $x_j=0$ then $\|P_{\thb_j}x_j - f_{j}(\textbf{x},\textbf{u},\betab)\|^2$ is not necessarily zero. Also, in this case the equations of the ODE are not independent. In a general setting, each $x_j$ is affected by $x_1,\dots,x_n$.  

To circumvent previous problem we follow an approach that reduces the nonhomogeneous system to an homogeneous one. The key aspect is to consider that each $f_j$ is a function of $\betab$ that does not depends directly on $\textbf{x}$ but that still reflects the dynamics of the system. To do so, we replace $\textbf{x}$ by some fixed surrogate, namely $\textbf{x}'(t)$, which is independent of $\thetab$ and $\betab$ and that it is estimated using the data. In section 5 we elaborate on the definition of an appropriate $\textbf{x}'$.

In general, in order to find a RKHS representation of the ODE system we assume that a Green's function $G_j$ of each $P_{\thetab_j}$ exists. Consider
\begin{equation}\label{eq:x.star}
\tilde{x}_j(t) = x_j(t) - x_j^*(t),
\end{equation}
where $x_j^*(t) =  \int_T G_j(z,t) f_j(\textbf{x}'(z),\textbf{u}(z),\betab) dz$ is a collection of solutions of the differential equation. Since $P_{\theta_j}$ is a linear operator we obtain that, for all $\tilde{x}_j$,
$$P_{\thetab_j}\tilde{x}_j(t) = P_{\theta_j}x_j(t) - P_{\theta_j}x_j^*(t) = P_{\theta_j}x_j(t) -f_j(\textbf{x}'(z),\textbf{u}(z),\betab),$$
which includes the trivial solution $\tilde{x}_j=0$. We can now write
\begin{equation}
\Omega(\tilde{x}_j) = \|P_{\theta_j}x_j(t) - P_{\theta_j}x_j^*(t)\|^2 = \|P_{\thetab_j}\tilde{x}_j\|^2,  
\end{equation}
which can be studied and computed as a norm for $\tilde{x}_j$ in $\mathcal{H}_{\thb_j}$ equivalent to (\ref{eq:norm}). Note that the surrogate $\textbf{x}'$ is essential to linearize the system and to write $\Omega(\tilde{x}_j)$ as a norm. In practice, noise samples are obtained for $x_j$ and not for $\tilde{x}_j$. Therefore to focus the inference problem on $\tilde{x}_j$ requires to transform the original data. Details are given in Section 5. 


\section{Approximate ODE inference}\label{sec:implementation}

The goal of this section is to provide computational details to infer the set of parameters $\{\thetab,\betab\}$ using the approximate ODE representation described in Section 4. As detailed in Section 4.1, a definition of each $x'_j$, a surrogate of $x_j$ is required. In this work we express each $x'_j$ in terms of a penalized splines basis expansion, 
\begin{equation}\label{eq.xprime}
x'_j(t) = \sum_{k=1}^qc_{jk}\phi_k(t) = \textbf{c}_j^T\phib(t), 
\end{equation}
for $\phib = (\phi_1(t),\dots\phi_{q}(t))$ and where each $\phi_j(t)$ a piecewise polynomial function of degree $k$ whose brake points or knots are located at $t_1,\dots,t_n$. The coefficients $\textbf{c}_j$ can be explicitly estimated by $\textbf{c}_j = (\Phi^T\textbf{W}\Phi)^{-1}\Phi^T\textbf{W}\textbf{y}_j$ for $\Phi$ the design matrix and where $\textbf{W}$ is a weight matrix which allows for possible covariance structure among the residuals See \citep{bib:ramsay97} for details. The number of basis $q$ is assumed to be large enough to capture the variation of the solutions of the ODE. Here we assume that coefficients $\textbf{c}_j$ are fixed values obtained by smoothing the data but further generalizations are possible.

\begin{definition}
Consider the penalized likelihood model in (\ref{def:like.univ.ode.pen}). Consider the objects $\textbf{t}$, $\textbf{y}_{j}$, $\textbf{x}_j$ and $\textbf{P}_{\thetab_j}$ as defined above. Let $\hat{x}'_1,\dots,\hat{x}'_m$ be estimates of (\ref{eq.xprime}) for $j=1,\dots,m$. We define the approximated pseudo-log-likelihood of the ODE model associated to (\ref{def:like.univ.ode.pen}) as
$$l_{\lambda}(\thetab,\betab | \textbf{y}(\textbf{t}),\hat{\textbf{x}}' )=- \sum\limits_{j=1}^{m}\frac{1}{2\sigma^2_j}\|\textbf{y}_{j}-\textbf{x}_j\|^2 - \frac{\lambda}{2} \sum\limits_{j=1}^{m} \| \textbf{P}_{\thb_j} \textbf{x}_j - f_{j}(\hat{\textbf{x}}'(\textbf{t}),\textbf{u}(\textbf{t}),\betab)\|^2$$
where $\lambda>0$ and $\hat{\textbf{x}}'(\textbf{t}) = (\hat{x}_{1}(\textbf{t}),\dots,\hat{x}_{m}(\textbf{t}))^T$.
\end{definition}
Next, we show how to compute $l_{\lambda}$ in practice.
\begin{proposition}\label{prop:Mstep}
Assume that $\textbf{P}_{\thetab_j}^{-1}$ exists and define,
\begin{equation}\label{eq:transform}
\tilde{\textbf{y}}_{j}= \textbf{y}_{j} - \textbf{P}_{\thetab_j}^{-1}f_j(\textbf{u}(\textbf{t}),\hat{\textbf{x}}'(\textbf{t}),\betab).
\end{equation}
 for $j=1,\dots,m$. Consider the function 
\begin{equation}
g_{\lambda}(\thetab,\betab | \textbf{y}(\textbf{t})) = \sum_{j=1}^m \frac{1}{2\sigma_j^2}  \tilde{\yb}_{j}^T\left[\Ib -  (\Ib + \sigma^2\lambda  \textbf{K}_{\thb_j}^{-1} )^{-1} \right]\tilde{\yb}_{j}.
\end{equation}
It holds that 
$$\{\hat{\thetab},\hat{\betab}\}=arg  \max_{\thetab,\betab} l_{\lambda}(\thetab,\betab, \textbf{y}(\textbf{t})) = arg  \max_{\thetab,\betab} g_{\lambda}(\thetab,\betab| \textbf{y}(\textbf{t})).$$
\end{proposition}
Remark that we do not need the explicit computation of $\textbf{K}_{\thb_j}$. Only its inverse is needed, which can be obtained directly from (\ref{eq:K}).  Notice that (\ref{eq:transform}) performs the data transformation equivalent to (\ref{eq:x.star}) that is needed to obtain a RKHS representation of the ODE in general cases. Optimization of (\ref{prop:Mstep}) with a conjugate gradient method produces estimates of $\{\thetab, \betab \}$. If the set of parameters of the systems is separable by equations, independent optimization can be done for those, which helps to avoid local minima and speed up the procedure. Finally, estimates for each $\hat{\textbf{x}}_j$ are available by means of
\begin{equation}\label{eq:alpha}
\hat{\textbf{x}}_j =  \textbf{K}_{\hat{\theta}_j}\hat{\alphab}_j +\textbf{P}_{\thetab_j}^{-1}f_j(\textbf{u}(\textbf{t}),\hat{\textbf{x}}'(\textbf{t}),\hat{\betab})
\end{equation}
where $\hat{\alphab}_j= ( \textbf{K}_{\hat{\theta}_j} + \lambda \sigma_j^2 \textbf{I} )^{-1}\tilde{\textbf{y}}_{j}$. 

\subsection{Model selection}
The effective number of parameters for each equations is defined as $df_j = Tr(\textbf{K}_{\hat{\theta}_j}( \textbf{K}_{\hat{\theta}_j} + \lambda \sigma_j^2 \textbf{I} ))^{-1}$ where $Tr(\cdot)$ represents the trace operator. We propose as model selection criteria for $\lambda$ the Akaike information criteria \citep{bib:Akaike1974}, which is defined in our context as
\begin{equation}\label{eq:AIC}
\mbox{AIC}(\lambda) = -2 l_{\lambda}(\thetab,\betab | \textbf{y}(\textbf{t}),\hat{\textbf{x}}' ) + 2 \sum_{j=1}^m df_j.
\end{equation}
In practice, the minimun AIC for a grid of penalization parameters $\lambda_1,\dots,\lambda_k$ should be used to select the optimal model. Note, however, that one can simply select a large value of $\lambda$ to force the regularization approach to find an exact solution of the ODE model.

\subsection{Variance of the parameter estimates and confidence intervals}
Confidence intervals and standard errors of the parameter estimates can be obtained from the Hessian matrix $\textbf{H}_{l_{\lambda}}(\hat{\thetab},\hat{\betab})$, which is available as output of the Conjugate Gradient algorithm \citep{Nocedal2006NO} used to optimize $\l_{\lambda}$. The covariance matrix of the parameter estimates is therefore obtained as $\hat{\Sigma} = -\textbf{H}_{l_{\lambda}}(\hat{\thetab},\hat{\betab}) ^{-1}$ and  its diagonal (variances of the parameters) used to estimate calculate the Wald confidence intervals.

\section{Examples using synthetically generated data}

\subsection{Explicit ODE versus regularization approach}

\begin{figure}[t!]
  \begin{center}
   \mbox{
     \subfigure[Negative log likelihood and the penalized log-likelihood of the model for one of the generated data sets.] {\includegraphics[height=6.8cm,width=6.8cm]{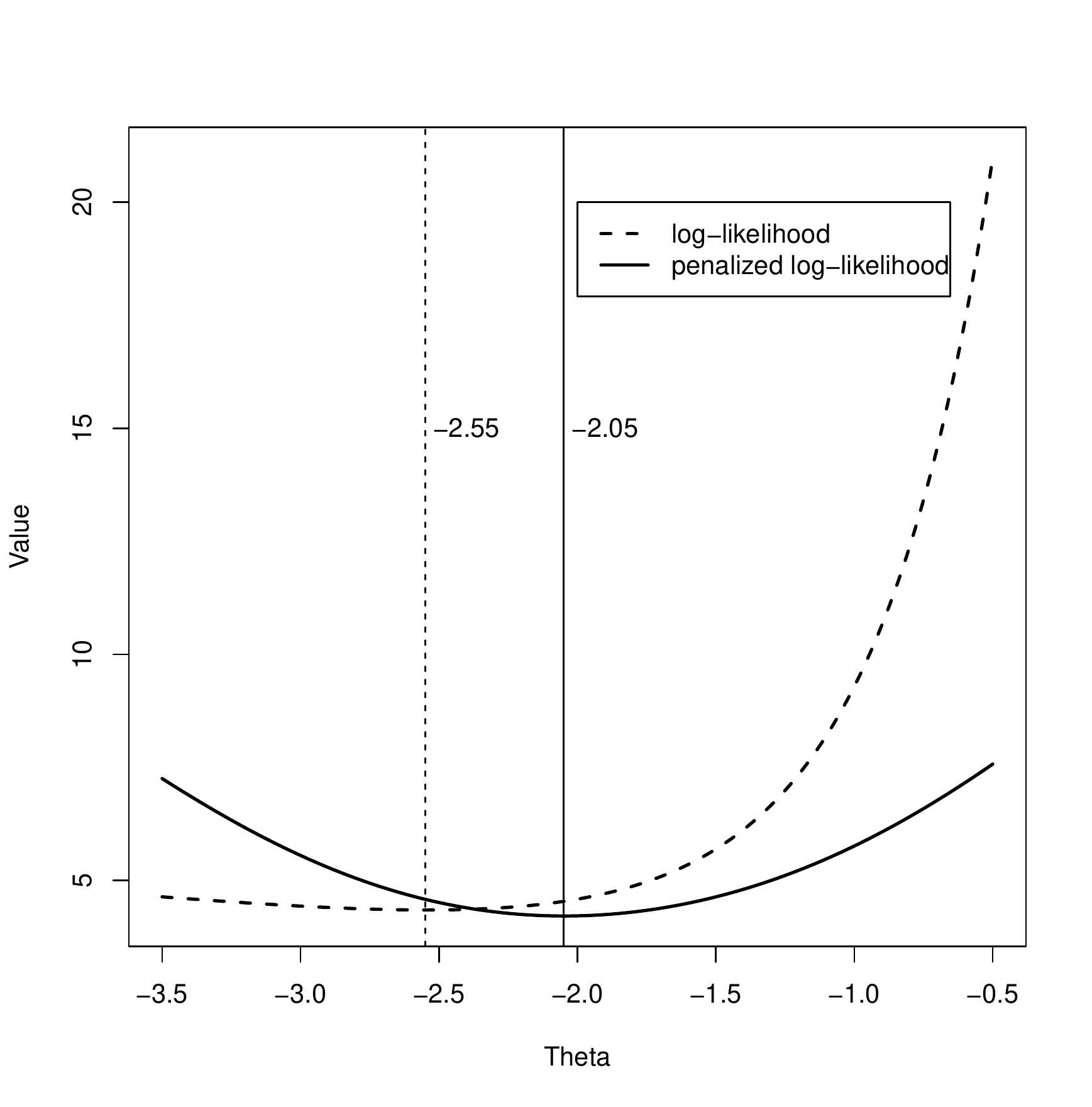}}\quad 
      \subfigure[Simulated data, true solution, and the two estimated solutions using the MLE and RKHS approaches.] {\includegraphics[height=6.8cm,width=6.8cm]{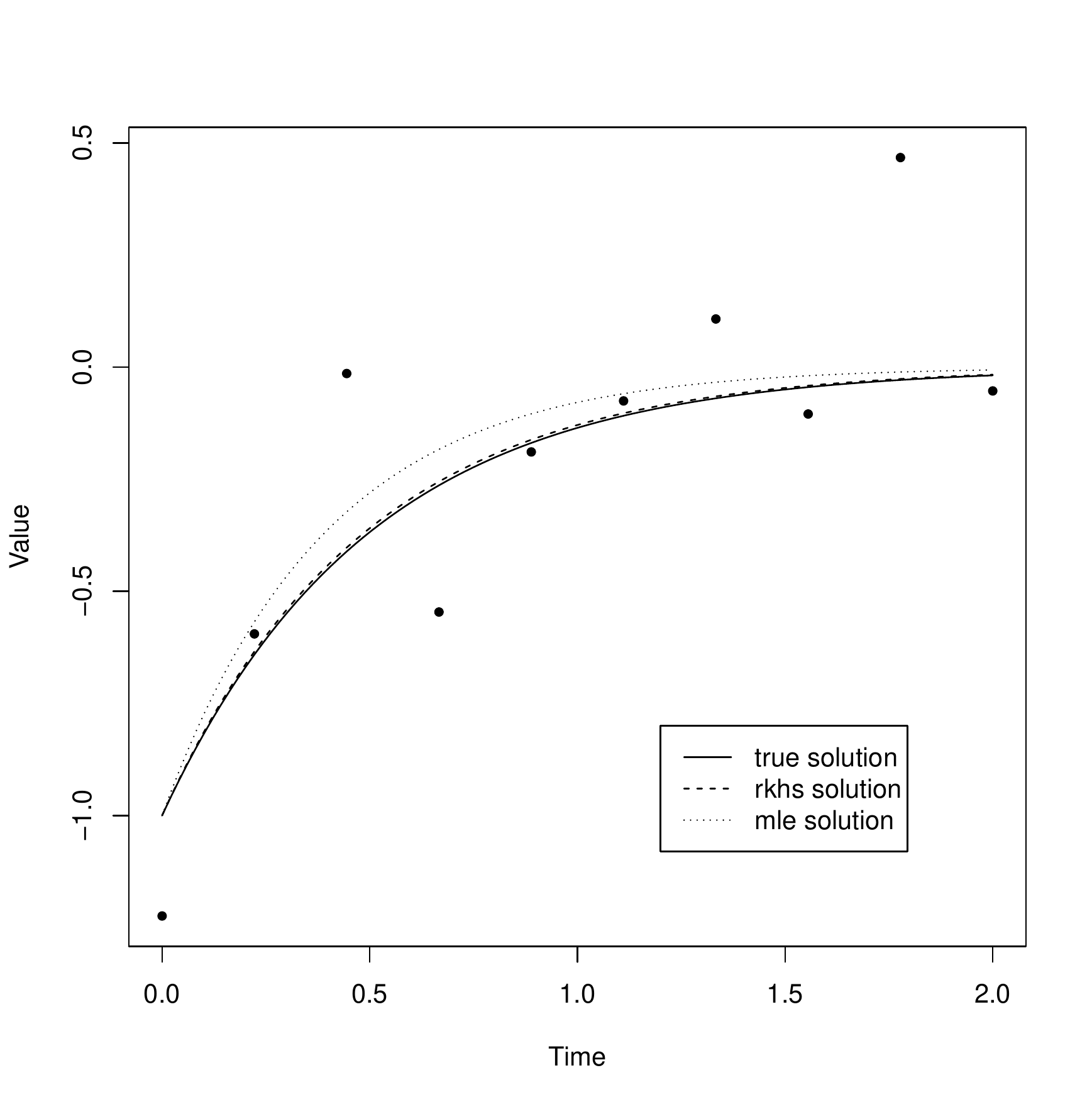}}
        }
 \end{center}
\caption{Results obtained for the differential equation  $dx/dt - \theta x =0$}\label{figure:example}
\end{figure}

In this section we use a toy example to illustrate the advantages of using a regularization approach to estimate the parameters of a dynamical system. We consider the differential equation  $dx/dt - \theta x =0$ where $D=d/dt$. For fixed $\theta$ and initial condition $x(0)$ the  solution of  the differential equation is given by $x(t) = x(0) \exp\{\theta t\}$. We fix $\theta=-2$, $x(0)=-1$ and we generate 500 samples of 10 equally spaced points in the interval $[0,2]$ using Gaussian noise with $\sigma=0.25$. For each sample we calculate the maximum likelihood estimator (MLE) of $\theta$ and our RKHS based estimator for $\lambda$ selected by means of the Akaike Information Criteria (AIC) (see Section 5.1). The averaged absolute deviance to the true parameter of the MLE is 0.73 with a standard deviation of 1.03 whereas the averaged error for the penalized approach is $0.53$ with a standard deviation of 0.38. In Figure \ref{figure:example} we show the results for one run of the experiment. Figure \ref{figure:example} a)  shows the negative log likelihood and the penalized log-likelihood of the model for one of the generated data sets. The penalized approach results in an 'improvement' of the original likelihood for parameter estimation with a minimum closer to the true value of the parameter. Also note that the original negative log likelihood becomes extremely flat for small values of the parameters, which can produce computational problems in the optimization step. Figure \ref{figure:example} b) shows the MLE and RKHS solutions together with the true function $x(t) = -\exp\{2t\}$ for $t\in [0,2]$. Penalizing the likelihood improves the estimator in this example. The true function $x$ is better approximated using the penalized approach due to the finite sample bias of the MLE. Also the estimate of the parameter is closer to the true value of $\theta$ in this particular realization ($\hat{\theta}_{MLE} =-2.55$ vs.  $\hat{\theta}_{RKHS} = -2.05$).

\subsection{Lotka volterra equations}
 \begin{figure}[t!]
  \begin{center}
   \mbox{
     \subfigure[True solutions and the data in the Lotka-Volterra experiment for $n=100$ and $\sigma=0.25$.] {\includegraphics[height=6.7cm,width=6.7cm]{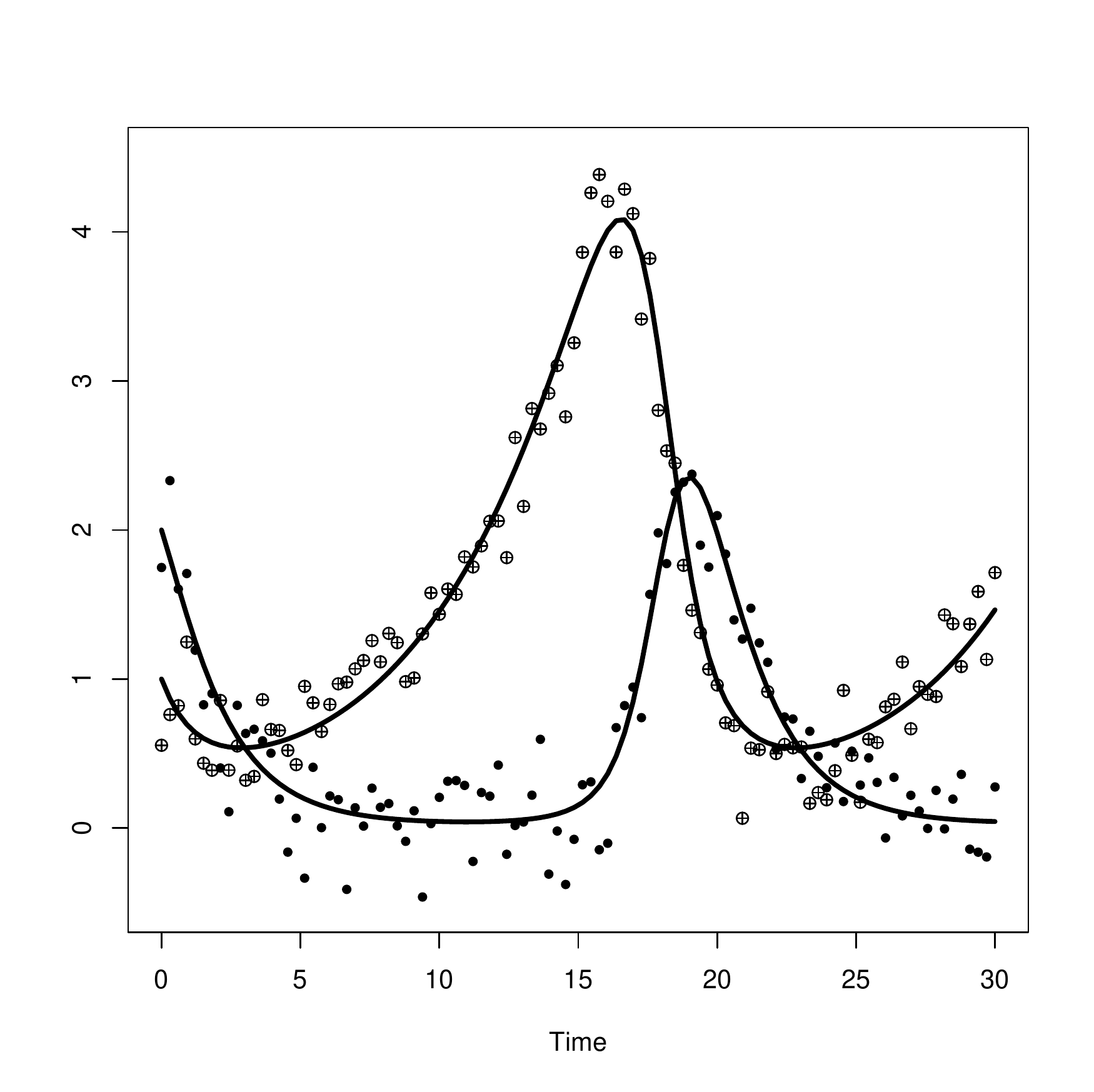}}\quad 
      \subfigure[Computational cost comparison between the explicit MLE approach and the proposed penalized RKHS based approach.] {\includegraphics[height=6.7cm,width=6.7cm]{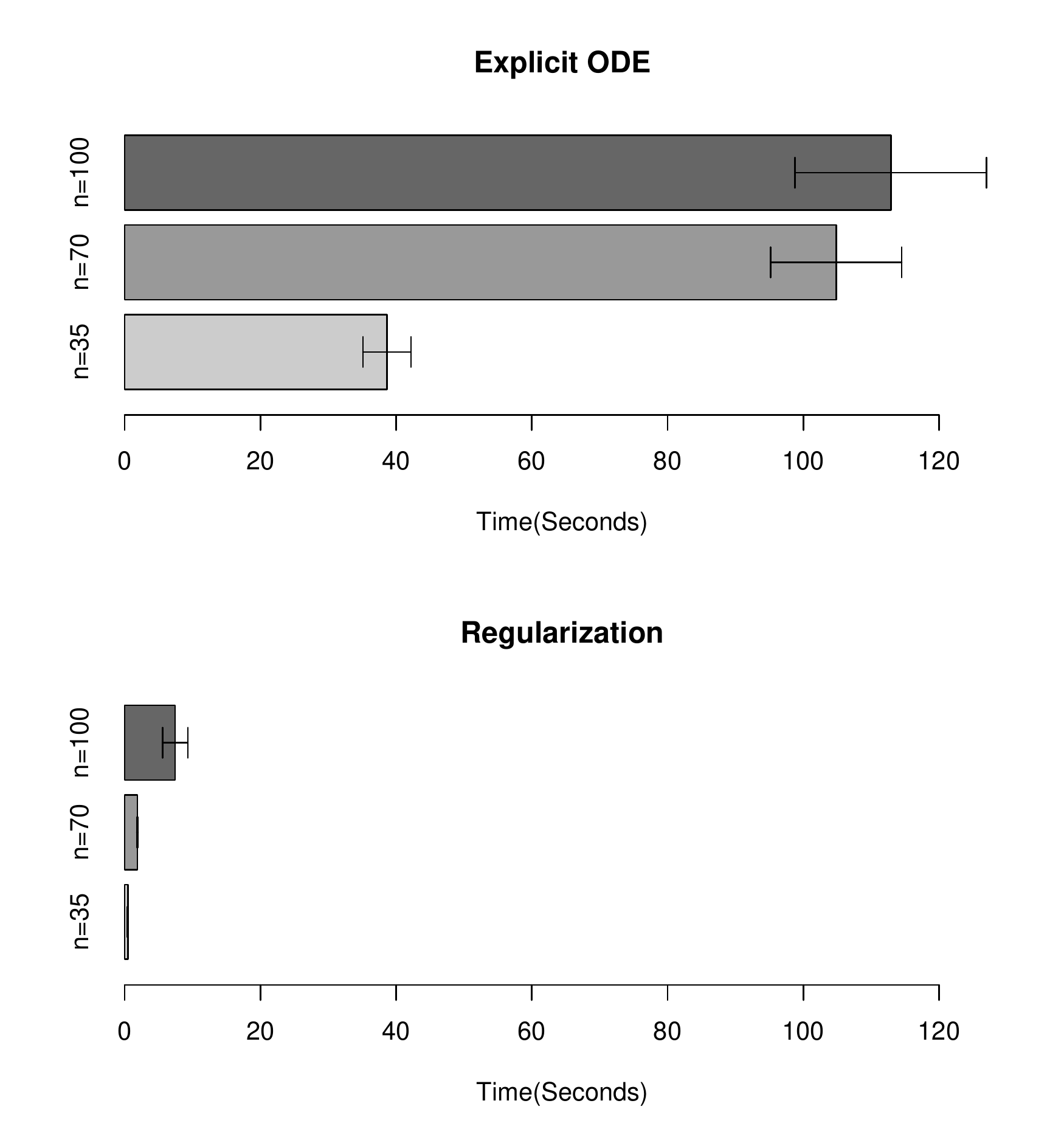}}
        }
 \end{center}
\vspace{-0.6cm}\caption{Results obtained for the  Lotka-Volterra equations. }\label{figure:LV}
\end{figure}

In this experiment we work with the Lotka-Volterra system of differential equations originally proposed in the theory of auto-catalytic chemical reaction \citep{Lotka1910}. The formulation of the system is given by
\begin{equation}\label{eq:LVmodel}
\frac{d x_1}{dt} = x_1(\theta_1 - \beta_1 x_2), \,\,\,\,\frac{dx_2}{dt}  = - x_2(\theta_1 - \beta_2 x_2)\\
\end{equation}
where $\thetab=(\theta_1, \theta_2)^T$, $\betab = (\beta_1, \beta_2)^T$ are the parameters.

Our aim is to evaluate the accuracy and speed of our RKHS penalized approach in comparison with the classical MLE approach. To do so, we run a simulation study for fixed $\theta_1=0.2 $, $\beta_1=0.35$ $\theta_2=0.7$, $\beta_2=0.40$ and initial conditions $x_{1}(0) = 1 $ and $x_{2}(0) = 2$. We generate samples made up of $n$ fixed and equally spaced independent noisy observations of the state variables $x_1$ and $x_2$ in the interval $T=[0,30]$ that we perturb with zero mean Gaussian noise with standard deviation $\sigma$. We generate data for the samples sizes $n=35,70,100$ and two noise scenarios $\sigma=0.1,0.25$. In Figure \ref{figure:LV} a) we show the true solutions of the model for the above mentioned parameters together with the data of one of the simulations.

In order to apply the proposed approach we obtain the functions $\hat{x}'_1$ and $\hat{x}'_2$ using penalized splines and for fix $\lambda=100$. To perform the MLE estimation we use an conjuate gradient algorith with 10 different initial 
values of the parameters (randomly generated in the interval $[0,1]$) and we use the likelihood value to select the best candidate. 

In Figure \ref{figure:LV} b) we show a computational time comparative for the averaged running times. 
The RKHS-based is 120.08, 24.06  and 14.41 times faster than the explicit ODE approach for $n=35, 70$ and $100$ respectively. In Table 1 we show the mean square errors of the
estimates with respect to the true parameters for 100 runs of the experiment. For $n=35$ the penalized RKHS approach performs significantly better than the explicit ODE estimates, which is explained by the empirical bias suffered by the MLE approach illustrated in Section 6.1. For $n=75,100$ both methods work similarly in terms of precision. Notice that the noise in the data is reflected in the precision of the estimates for both techniques; the errors are larger in all cases for $\sigma=0.25$.

\begin{table}[t!]
\footnotesize {
\begin{center}\caption{Mean square error for the inferred parameters in the Lotka-Volterra model. Standard deviations shown in parenthesis. The true value of the parameters are fixed to $\theta_1=0.2 $, $\beta_1=0.35$ $\theta_2=0.7$, $\beta_2=0.40$}
\begin{tabular}{|c|c|c|c|c|c|c|}
\hline
\multicolumn{7}{ |c| }{Lotka-Volterra ODE model} \\
\hline
$\sigma$ &n & Method& $|\theta_1 - \hat{\theta}_1|$ & $|\beta_1 - \hat{\beta}_1|$ & $|\theta_2 - \hat{\theta}_2|$ & $|\beta_2 - \hat{\beta}_2|$ \\
   \hline
\multirow{6}{*}{$0.1$}&\multirow{2}{*}{35} & RKHS     & 0.0002 (0.0003) & 0.0007  (0.0007) & 0.0031  (0.0036) & 0.0014  (0.0014) \\ 
&			& MLE   									& 0.0016  (0.0088) & 0.0063 (0.0425) & 0.0422  (0.1809) & 0.0227  (0.1064) \\[0.15cm]
&\multirow{2}{*}{70} & RKHS   						& 0.0001  (0.0001) & 0.0002  (0.0002) & 0.0009  (0.0011) & 0.0003  (0.0004) \\ 
&			& MLE   									& 0.0000  (0.0001) & 0.0001  (0.0006) & 0.0017  (0.0034) & 0.0005  (0.0010) \\ [0.15cm]
&\multirow{2}{*}{100}  & RKHS   						 & 0.0001  (0.0001) & 0.0001  (0.0002) & 0.0005  (0.0006) & 0.0002 (0.0002) \\ 
&			& MLE   
& 0.0000  (0.0001) & 0.0002  (0.0010) & 0.0013  (0.0023) & 0.0004  (0.0008) \\ 
   \hline   \hline
\multirow{6}{*}{$0.25$}&\multirow{2}{*}{35}     	& RKHS   & 0.0010  (0.0013) & 0.0017  (0.0024) & 0.0111  (0.0205) & 0.0038 (0.0059) \\ 
&			& MLE   										& 0.0029  (0.0180) & 0.0081  (0.0392) & 0.0173  (0.0487) & 0.0078  (0.0359) \\ [0.15cm]
&\multirow{2}{*}{70} & RKHS  							& 0.0004 (0.0006) & 0.0008  (0.0009) & 0.0042  (0.0047) & 0.0015  (0.0019) \\ 
&			& MLE    									& 0.0007  (0.0025) & 0.0030  (0.0115) & 0.0151  (0.0474) & 0.0062  (0.0301) \\ [0.15cm]
&\multirow{2}{*}{100}    & RKHS 							& 0.0003  (0.0004) & 0.0005  (0.0006) & 0.0034  (0.0043) & 0.0011  (0.0016) \\ 
&			& MLE  										& 0.0008  (0.0032) & 0.0028  (0.0116) & 0.0174  (0.0603) & 0.0083  (0.0387) \\  
   \hline   \hline
\end{tabular}\label{table:results}
\end{center}}
\end{table}

\section{Real example: Reconstruction of Transcription Factor activities in Streptomyces coelicolor}

A gene regulatory network consists of a gene encoding a transcription factor (TF) together with the genes it regulates (genes whose activity can be activated or repressed by binding to the DNA). 
In the absence of reliable technology to measure the activity of the TF (number of TF-protein molecules in the cell), the problem is to reconstruct it from gene expression data of its target genes.

 \begin{figure}[t!]
  \begin{center}
   \mbox{
     \subfigure[Gene SCO3235, reconstructed profile. Circles and crosses represent the observed data and lines the obtained profiles. The estimated variance and initial conditions are $\hat{\sigma}^2=0.016$ and $\hat{x}(0)=-0.39$. The estimated parameters for this gene are $\beta_{1}$ = 0.65 (stdev= 0.57), $\beta_{2} = 1.07 (0.51)$, $\beta_{3}$ = 2.09 (0.26) and $\theta$=1.06 (0.21).] {\includegraphics[height=6.8cm,width=6.8cm]{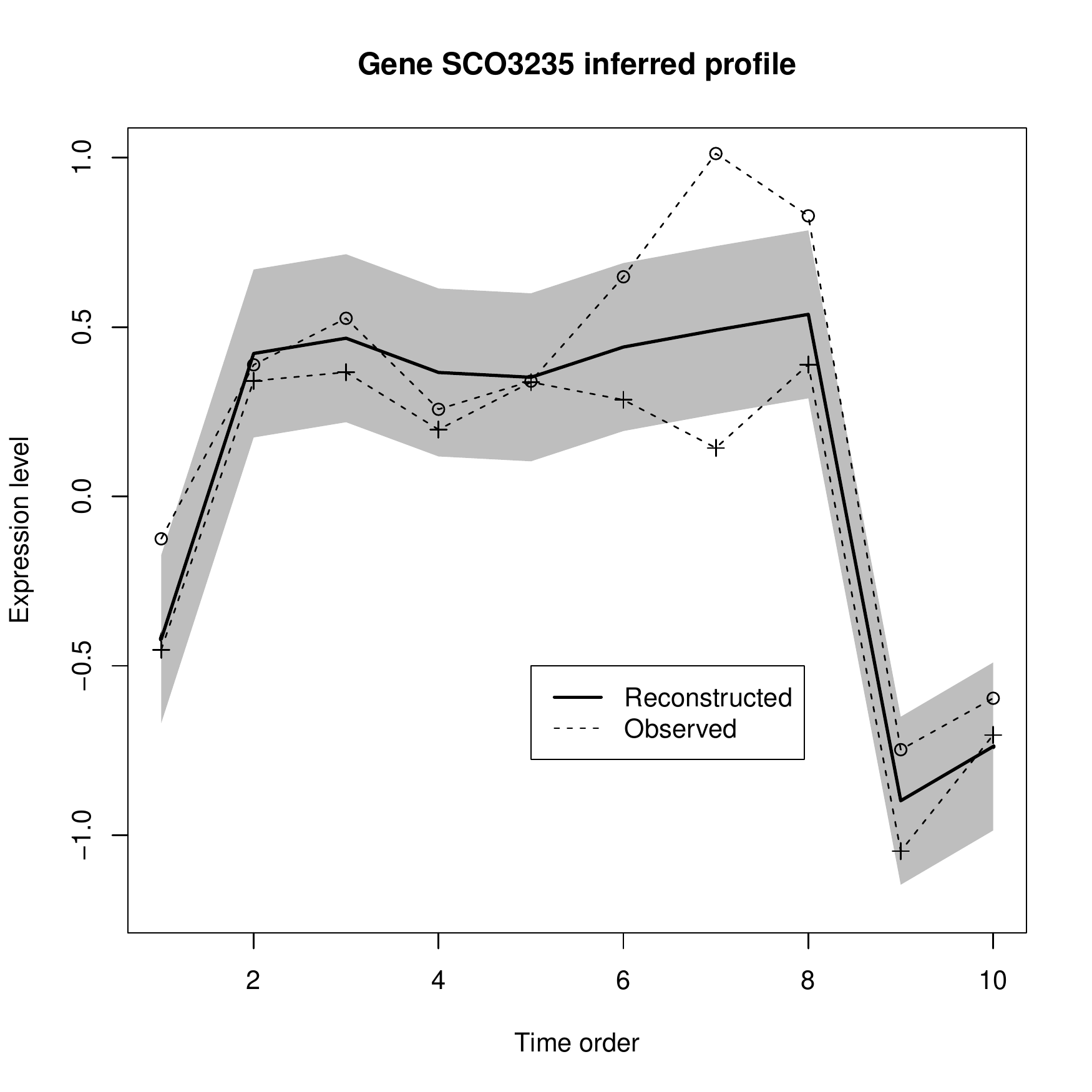}}\quad 
      \subfigure[Reconstructed activity of the master activator cdaR scaled between 0 an 1. Circles and crosses represent the data obtained in two independent experiments not used in the estimation process.] {\includegraphics[height=6.8cm,width=6.8cm]{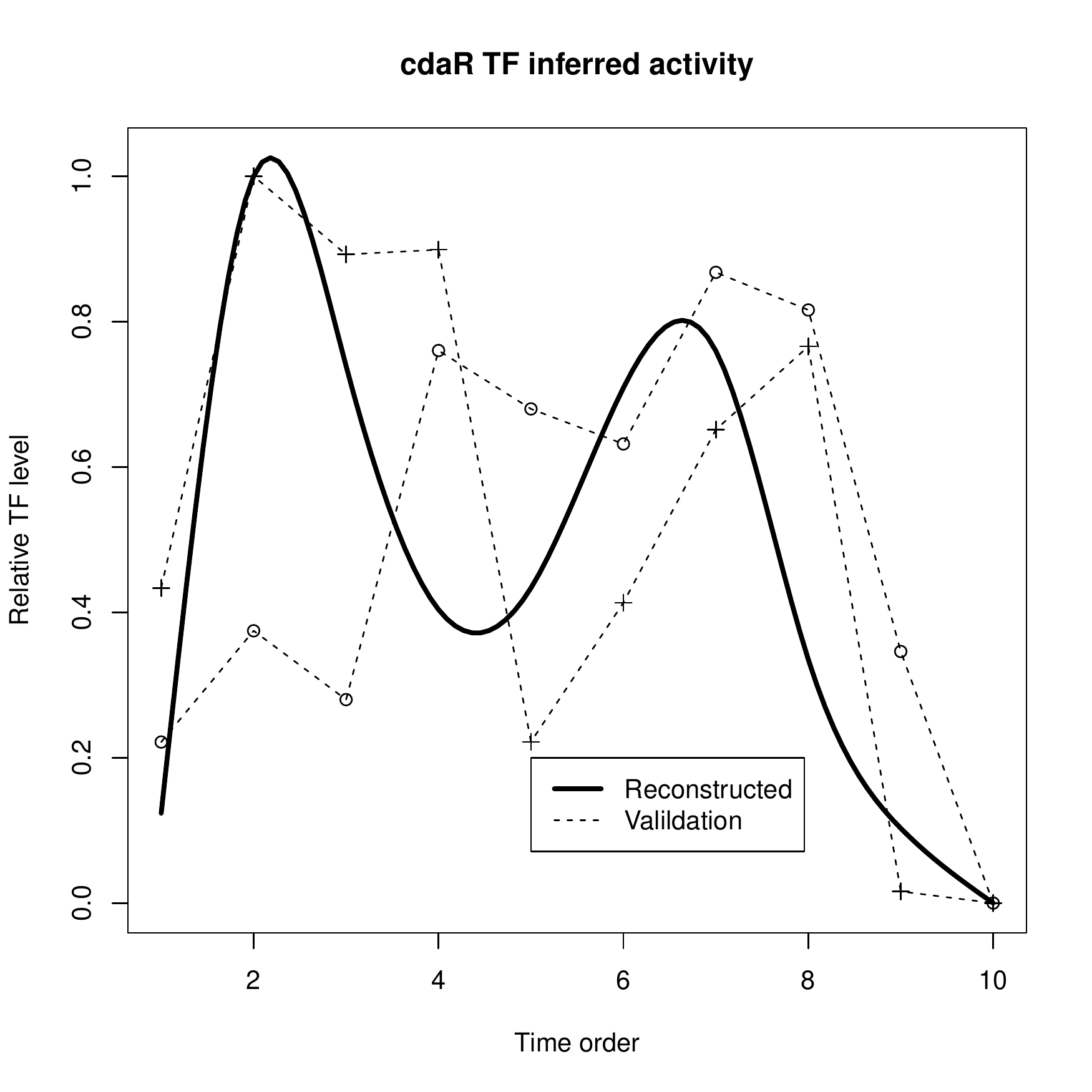}}
        }
 \end{center}
\vspace{-0.6cm}
 \caption{Reconstructed genes profiles and master activator cdaR.}\label{figure:rec.profiles}
\end{figure}

In this experiment we work with a data set of genes expression levels in the \emph{Streptomyces coelicolor} bacterium. The goal is to reconstruct the activity of the transcription factor (TF) \emph{cdaR} using 
10-points time-series gene expression data of 17 genes. For each gene, two different series corresponding to a wild type and mutant type bacterium (for which a transcriptional regulator 
cdaR has been knocked out) are available. Measurements are available at time points (in mins.) $\textbf{t}=\{16,18, 20,21,22,23,24,25,39,67\}$. The 
importance of understanding the behaviour of the cdaR relies on the fact that it is partially responsible for the production of a particular type of antibiotic.

Following \citet{Khanin07} we assume that changes in the expression levels of the genes are caused by changes in the cdaR protein and the mRNA degradation. We denote by $\eta(t)$  the activity profile of the regulator cdaR at time $t$ and by $x_j(t)$ the expression level of each gene $j$ in time $t$. This regulatory system is modelled by
\begin{equation}\label{eq:systemEcoli}
\frac{d}{dt}x_j(t) + \theta_j x_j(t)  = \beta_{1j} + \beta_{2j} \frac{\eta(t)}{\beta_{3j}+ \eta(t)}, 
\end{equation}
where $\theta_j$ is the rate of mRNA degradation, $\beta_{2j}$ and $\beta_{3j}$ are gene-specific kinetic parameters for the gene $j$ and $\beta_{1j}$ is an additive constant that accounts for the basal level of transcription and the nuisance effects from micro-arrays. The goal is to use the available sample to reconstruct the levels of the activator $\eta(t)$, which is unobserved, and the gene profiles via the estimation of the parameters in  (\ref{eq:systemEcoli}). To do so we apply the procedure described in Sections 3 and 4. We assume that the variance is equal for all the genes. For each gene, we work with the average of the two available time series. We model the 
activator $\eta$ using a basis of cubic splines with equally spaced nodes, that is $\eta(t)=\sum_{j=1}^{15}a_j\phi_j$ where the $\phi_j$'s are elements of the basis and $a_1,\dots,a_{15}$ parameters to estimate. We apply the procedure described in Section 5. We select the optimal penalization parameter $\lambda$ by using the AIC  as model selection criterion. 

Figure \ref{figure:rec.profiles} a)  shows the estimated profile for the gene SCO3235, which fits well the observed data. The reconstructed genes profiles exhibit a similar fit for the remaining genes. The reconstructed cdaR activator is shown in Figure \ref{figure:rec.profiles} b) together two independent replicates profiles obtained from a different experiment and that were not used in the estimation process. The values are normalized between 0 and 1 since the activity of the cdaR protein is expressed in arbitrary units and can be interpreted as relative levels. The estimated profile fits the observed data showing two hills around time points 4 and 9 similarly to the genes profiles. This agrees with the fact that cdaR is an activator of the genes activity. These result shows the ability of the proposed approach to identify correctly unknown elements of the ODE systems through a proper estimation of the parameters of the model. The estimation of the parameters of this system takes around 15 mins in a personal laptop. Further estimates of the baseline MLE estiamators are available in  \citet{Khanin07}.

\section{Conclusions and discussion}

We have proposed a new method to estimate general systems of ordinary differential equations measured with noise. Our proposal is based on the penalization
of the likelihood of the problem by means of the ODE. A reproducing
kernel Hilbert space approach has provided the theoretical framework to make this idea
feasible. The concept of Green's function and the connection between linear differential operators
and Mercer kernels have been used to rewrite the penalized likelihood of the problem in a 
particular manner easy to handle in practice. 

The main merit of the method is its ability to perform in a single step the estimation problem without
solving the differential equation. In practice, our proposal is specially competitive in small sample scenarios as it is shown via simulation. A real example in system biology has been used to illustrate the utility of the method in scenarios with hidden components.

In the future, we aim to focus on futher Bayesian extensions of this work, the estimation of the regularization parameter $\lambda$ and on the analysis of the theoretical properties of the proposed method.





\bibliographystyle{elsarticle-harv}
\bibliography{ode2}

\begin{thebibliography}{23}
\expandafter\ifx\csname natexlab\endcsname\relax\def\natexlab#1{#1}\fi
\expandafter\ifx\csname url\endcsname\relax
  \def\url#1{\texttt{#1}}\fi
\expandafter\ifx\csname urlprefix\endcsname\relax\def\urlprefix{URL }\fi

\bibitem[{Akaike(1974)}]{bib:Akaike1974}
Akaike, H., 1974. A new look at the statistical model identification. IEEE
  Transactions on Automatic Control 19(6), 716--723.

\bibitem[{Aronszajn(1951)}]{bib:aroszajn51}
Aronszajn, N., 1951. Green's functions and reproducing kernels. Proceedings of
  the Simposium on Spectral Theory and Differential Problems, Oklahoma,
  355--411.

\bibitem[{Aroszajn(1950)}]{bib:aroszajn50}
Aroszajn, N., 1950. Theory of reproducing kernels. Transactions of the American
  Mathematical Society 68(3), 337--404.

\bibitem[{Berlinet and Thomas-Agnan(2005)}]{bib:Berlinet2004}
Berlinet, A., Thomas-Agnan, C., 2005. Reproducing kernel hilbert spaces in
  probability and statistics. Srpinger.

\bibitem[{Calderhead et~al.(2008)Calderhead, Girolami, and
  Lawrence}]{Calderhead08}
Calderhead, B., Girolami, M., Lawrence, N., 2008. Accelerating bayesian
  inference over nonlinear differential equations with gaussian processes. In:
  Neural Information Processing Systems. Vol.~22.

\bibitem[{{Chkrebtii} et~al.(2013){Chkrebtii}, {Campbell}, {Girolami}, and
  {Calderhead}}]{2013arXiv1306.2365C}
{Chkrebtii}, O., {Campbell}, D.~A., {Girolami}, M.~A., {Calderhead}, B., Jun.
  2013. {Bayesian Uncertainty Quantification for Differential Equations}. ArXiv
  e-prints.

\bibitem[{Cucker and Smale(2001)}]{bib:cucker2001}
Cucker, F., Smale, S., 2001. On the mathematical foundations of learning.
  Bulletin of the American Mathematical Society 39(1), 1--49.

\bibitem[{Duffy(2001)}]{duffy2001}
Duffy, D.~G., 2001. Green's Functions with Applications. Chapman and Hall/CRC.

\bibitem[{Green and Silverman(1994)}]{Green_Silverman_1994}
Green, P.~J., Silverman, B.~W., 1994. Nonparametric Regression and Generalized
  Linear Models: A Roughness Penalty Approach. Vol.~58. Chapman \& Hall.

\bibitem[{Hofmann et~al.(2008)Hofmann, Sch\"olkopf, and J.}]{Hofmann2008}
Hofmann, T., Sch\"olkopf, B., J., S.~A., 2008. Kernel methods in machine
  learning. Annals of Statistics 36, 1171--1220.

\bibitem[{Khanin et~al.(2007)Khanin, Vinciotti, Mersinias, Smith, and
  Wit}]{Khanin07}
Khanin, R., Vinciotti, V., Mersinias, V., Smith, C., Wit, E., 2007. Statistical
  reconstruction of transcription factor activity using michaelis�menten
  kinetics 63~(3), 816�823.

\bibitem[{Kimeldorf and Wahba(1970)}]{Kimeldorf1970}
Kimeldorf, G.~S., Wahba, G., 1970. A correspondence between bayesian estimation
  on stochastic processes and smoothing by splines. The Annals of Mathematical
  Statistics 41(2), 495--502.

\bibitem[{Lotka(1910)}]{Lotka1910}
Lotka, A.~J., 1910. Contribution to the theory of periodic reaction. J. Phys.
  Chem. 14~(3), 271–274.

\bibitem[{Ma et~al.(2010)Ma, Dai, Klein, Klein, Lee, and
  Wahba}]{Ma_Dai_Klein_Klein_Lee_Wahba_2010}
Ma, X., Dai, B., Klein, R., Klein, B. E.~K., Lee, K.~E., Wahba, G., 2010.
  Penalized likelihood regression in reproducing kernel hilbert spaces with
  randomized covariate data. Statistics 17~(1159), 46.

\bibitem[{Moguerza and Mu\~{n}oz(2006)}]{bib:moguerza06}
Moguerza, J.~M., Mu\~{n}oz, A., 2006. Support vector machines with
  applications. Statistical Science 21(4), 322--336.

\bibitem[{Nocedal and Wright(2006)}]{Nocedal2006NO}
Nocedal, J., Wright, S.~J., 2006. Numerical Optimization, 2nd Edition.
  Springer, New York.

\bibitem[{Parzen(1970)}]{Parzen1970}
Parzen, E., 1970. Statistical inference on time series by rkhs methods. In
  Proceedings 12th Biennial Seminar (R. Pyke, ed.) Canadian Mathematical
  Congress, Montreal., 1--37.

\bibitem[{Poggio and Girosi(1990)}]{58326}
Poggio, T., Girosi, F., sep 1990. Networks for approximation and learning.
  Proceedings of the IEEE 78~(9), 1481 --1497.

\bibitem[{Ramsay et~al.(2007)Ramsay, Hooker, Campbell, and Cao}]{Ramsay07}
Ramsay, J.~O., Hooker, G., Campbell, D., Cao, J., 2007. Parameter estimation
  for differential equations: a generalized smoothing approach. Journal of the
  Royal Statistical Society: Series B (Statistical Methodology) 69~(5),
  741--796.

\bibitem[{Ramsay and Silverman(2006)}]{bib:ramsay97}
Ramsay, J.~O., Silverman, B.~W., 2006. Functional data analysis. Springer, New
  York, 2nd ed.

\bibitem[{Steinke and Scholkopf(2008)}]{Steinke08}
Steinke, F., Scholkopf, B., 2008. Kernels, regularization and differential
  equations. Pattern Recognition 41~(11), 3271--3286.

\bibitem[{Vapnik(1995)}]{bib:vapnik95}
Vapnik, V., 1995. The nature of statistical learning theory. Springer, New
  York.

\bibitem[{Wahba(1990)}]{bib:wahba90}
Wahba, G., 1990. Spline models for observational data. Series in Applied
  Mathematics, SIAM. Philadelphia.

\end{thebibliography}

\appendix
\section{}
\Skip
\noindent
\emph{\textbf{Proof of Proposition 2}}
\Skip
\noindent
We prove the proposition by showing that $l_{\lambda}(\thetab,\betab|\textbf{y}(\textbf{t}),\hat{\textbf{x}}' )$ and $g_{\lambda}(\thetab,\betab| \textbf{y}(\textbf{t}),\hat{\textbf{x}}' )$ are the same function. 
Consider the function
$$l_{\lambda}(\thetab,\betab | \textbf{y}(\textbf{t}),\hat{\textbf{x}}' )=- \sum\limits_{j=1}^{m}\frac{1}{2\sigma^2_j}\|\textbf{y}_{j}-\textbf{x}_j\|^2 - \frac{\lambda}{2} \sum\limits_{j=1}^{m} \| \textbf{P}_{\thb_j} \textbf{x}_j - f_{j}(\hat{\textbf{x}}'(\textbf{t}),\textbf{u}(\textbf{t}),\betab)\|^2.$$
Denote by $f_j = f_j(\textbf{u}(\textbf{t}),\hat{\textbf{x}}'(\textbf{t}),\betab)$. We can rewrite the first term as

\begin{eqnarray} \nonumber
\sum\limits_{j=1}^{m}\frac{1}{2\sigma^2_j}\|\textbf{y}_{j}-\textbf{x}_j\|^2 & = &\sum\limits_{j=1}^{m}\frac{1}{2\sigma^2_j}\|\textbf{y}_{j}-\textbf{P}_{\thetab_j}^{-1}f_j+\textbf{P}_{\thetab_j}^{-1}f_j -\textbf{x}_j\|^2\\ \nonumber
&= & \sum\limits_{j=1}^{m}\frac{1}{2\sigma^2_j}\|\tilde{\textbf{y}}_{j}-\tilde{\textbf{x}}_j\|^2, 
\end{eqnarray}
where  $\tilde{\textbf{y}}_{j}=\textbf{y}_{j}-\textbf{P}_{\thetab_j}^{-1}f_j$ and $\tilde{\textbf{x}}_j= \textbf{x}_j -\textbf{P}_{\thetab_j}^{-1}f_j $. 
Regarding the second term, we have that
\begin{eqnarray} \nonumber
\sum\limits_{j=1}^{m} \| \textbf{P}_{\thb_j} \textbf{x}_j - f_{j}\|^2 & = &\sum\limits_{j=1}^{m} \| \textbf{P}_{\thb_j} \textbf{x}_j - \textbf{P}_{\thb_j}\textbf{P}_{\thb_j}^{-1} f_{j}\|^2 \\ \nonumber
& = & \sum\limits_{j=1}^{m} \| \textbf{P}_{\thb_j}( \textbf{x}_j - \textbf{P}_{\thb_j}^{-1} f_{j})\|^2\\  \nonumber
& = & \sum\limits_{j=1}^{m} \| \textbf{P}_{\thb_j}\tilde{\textbf{x}}_j\|^2. \nonumber
\end{eqnarray}
Therefore, we can rewrite $l_{\lambda}(\thetab,\betab | \textbf{y}(\textbf{t}),\hat{\textbf{x}}' )$ as
$$l_{\lambda}(\thetab,\betab | \textbf{y}(\textbf{t}),\hat{\textbf{x}}' )=- \sum\limits_{j=1}^{m}\frac{1}{2\sigma^2_j}\|\tilde{\textbf{y}}_{j}-\tilde{\textbf{x}}_j\|^2 - \frac{\lambda}{2} \sum\limits_{j=1}^{m} \| \textbf{P}_{\thb_j} \tilde{\textbf{x}}_j \|^2.$$
Using the properties of RKHSs detailed in Section 4, we can write
$$l_{\lambda}(\thetab,\betab | \textbf{y}(\textbf{t}),\hat{\textbf{x}}' )=- \sum\limits_{j=1}^{m}\frac{1}{2\sigma^2_j}\|\tilde{\textbf{y}}_{j}-\Kb_{\thb_j}\alphab_j\|^2 - \frac{\lambda}{2} \sum\limits_{j=1}^{m} \alphab_j^T\Kb_{\thb_j}\alphab_j,$$
and expanding the first terms we obtain
$$- \sum_{j=1}^{m}\frac{1}{2\sigma_j^2}\biggl[\tilde{\textbf{y}}_{j}^T\tilde{\textbf{y}}_{j}-2\tilde{\textbf{y}}_{j}^T\Kb_{\thb_j}\alphab_j+\alphab_j^T\Kb_{\thb_j}^T\Kb_{\thb_j}\alphab_j+\sigma^2\lambda\alphab_j^T\Kb_{\thb_j}\alphab_j \biggr].$$
For fixed $\{\thetab, \betab\}$ and $\sigma_j^2$ the maximum of $l_{\lambda}(\thetab,\betab | \textbf{y}(\textbf{t}),\hat{\textbf{x}}' )$ is given for the vectors $\alphab_j=(\Kb_{\thb_j}+\sigma_j^2\lambda\Ib)^{-1}\tilde{\textbf{y}}_{j}$. We set the derivative to zero to find the point of maximum. Substituting each $\alphab_j$ and simplifying we obtain that
\begin{eqnarray}\nonumber
l_{\lambda}(\thetab,\betab | \textbf{y}(\textbf{t}),\hat{\textbf{x}}' )  & =   & 
  -\sum\limits_{j=1}^{m}\frac{1}{2\sigma_j^2} \biggl [ \tilde{\textbf{y}}_{j}^T\tilde{\textbf{y}}_{j}-2\tilde{\textbf{y}}_{j}^T\Kb_{\thb_j}(\Kb_{\thb_j}+\sigma_j^2\lambda\Ib)^{-1}\tilde{\textbf{y}}_{j} \\ \nonumber
&+& \tilde{\textbf{y}}_{j}^T(\Kb_{\thb_j}+\sigma_j^2\lambda\Ib)^{-1}(\Kb_{\thb_j}+\sigma_j^2\lambda\Ib)\Kb_{\thb_j}(\Kb_{\thb_j}+\sigma^2\lambda\Ib)^{-1}\tilde{\textbf{y}}_{j} \biggl ]  \\ \nonumber
    & =   & - \sum\limits_{j=1}^{m}\frac{1}{2\sigma^2}\biggl[\tilde{\textbf{y}}_{j}^T\tilde{\textbf{y}}_{j}-\tilde{\textbf{y}}_{j}^T\Kb_{\thb_j}(\Kb_{\thb_j}+\sigma_j\lambda\Ib)^{-1}\tilde{\textbf{y}}_{j}\biggr] \\ \nonumber
    &= &-\sum\limits_{j=1}^{m}\frac{1}{2\sigma_j^2}\tilde{\textbf{y}}_{j}^T\biggl[\Ib- (\Ib+\sigma_j^2\lambda  \textbf{K}_{\thb_j}^{-1})^{-1} \biggr]\tilde{\textbf{y}}_{j} \\
   &= & g_{\lambda}(\thetab,\betab| \textbf{y}(\textbf{t}),\hat{\textbf{x}}' )
 \nonumber
\end{eqnarray}
as we aimed to prove.







\end{document}